\documentclass[aps,prd,preprint,amsmath,amssymb,showpacs]{revtex4}
\usepackage{epsfig}
\def\ph{\phantom}
\def\beq{\begin{equation}}
\def\eeq{\end{equation}}
\def\hlf{\frac{1}{2}}
\def\del{\triangle}
\def\over{\overline}
\begin{document}
\title{Radiation Damping in Einstein--Aether Theory}
\author{Brendan Z. Foster}
\email[]{bzf@umd.edu} \affiliation{Department of Physics,
University of Maryland, College Park, MD 20742-4111, USA}
\date{September 23, 2008}
%
%
\begin{abstract}
This work concerns the loss of energy of a material system due to
gravitational radiation in Einstein--aether theory---an
alternative theory of gravity in which the metric couples to a
dynamical, timelike, unit-norm vector field.  Derived to lowest
post-Newtonian order are waveforms for the metric and vector
fields far from a nearly Newtonian system and the rate of energy
radiated by the system.  The expressions depend on the quadrupole
moment of the source, as in standard general relativity, but also
contain monopolar and dipolar terms. There exists a one-parameter
family of Einstein--aether theories for which only the quadrupolar
contribution is present, and for which the expression for the
damping rate is identical to that of general relativity to the
order worked to here.  This family cannot yet be declared
observationally viable, since effects due to the strong internal
fields of bodies in the actual systems used to test the damping
rate are not included.
\end{abstract}
\maketitle
%
%
\section{Introduction}
\label{INTRO}
That the world exhibits exact Lorentz invariance is a central
hypothesis of modern physics, and as such, warrants scrutiny.
Interest in testing this hypothesis has been growing in recent
years.  This is likely due to hints that this symmetry may be
broken in various candidates for theories of quantum gravity and
models of physics beyond the Planck scale, including
strings~\cite{Kostelecky:1988zi}, loop quantum
gravity~\cite{Gambini:1998it}, and noncommutative field
theories~\cite{Hewett:2000zp}. The review~\cite{Mattingly:2005re}
discusses various theoretical models that feature Lorentz-symmetry
violating effects and observational searches for violations.

If we wish to incorporate Lorentz violation into a gravitational
setting, we must employ a mechanism that breaks this symmetry
while preserving the distinct symmetry of diffeomorphism
invariance. Einstein--aether theory is a variation of general
relativity (GR) that does just this. Einstein--aether theory, or
`ae-theory' for short, is a classical, vector-tensor theory of
gravity in which the vector field `aether' is constrained to be
everywhere timelike and of fixed norm. The aether can be thought
of as an effective remnant of unknown, Planck scale,
Lorentz-violating physics.  It defines a ``preferred" frame, while
its status as a dynamical field preserves diffeomorphism
invariance.  The condition on the vector norm (which can always be
scaled to unity) ensures that the aether just picks out a
preferred direction and removes instabilities in the unconstrained
theory (see~\cite{Elliott:2005va}). A review of properties of this
theory and references to earlier work, can be found
in~\cite{Eling:2004dk}.

Much work on ae-theory has focused on deriving observational
constraints on the theory's free parameters, denoted here as $c_n;
(n=1,\dotsc,4)$.  Recent efforts have determined the form of the
theory's post-Newtonian expansion and the nature of linear, source
free wave phenomena, and have demonstrated that the theory remains
observationally healthy for a large range of $c_n$ values. The
parameterized post-Newtonian (PPN)
expansion~\cite{Will:1993ns,Will:2001mx} describes weak field,
slow motion
 gravitational effects in terms of ten parameters
that depend on the theory's free parameters.
 The PPN parameters have been constrained by solar system based tests
to have the values predicted by GR, up to small errors.  In
ae-theory, only two of the ten parameters, $\alpha_1$ and
$\alpha_2$, differ from the GR
values~\cite{Eling:2003rd,Graesser:2005bg,Foster:2005dk}. These
can be set to their GR values of zero by special choice of two of
the four $c_n$. The propagation of plane waves on a flat metric,
constant aether background was examined in~\cite{Jacobson:2004ts}.
It was shown that there generally exist five independent wave
modes---two spin-2, two spin-1, and one spin-0---that travel at
three different speeds that depend on the $c_n$.  The energy
densities associated with these modes were derived
in~\cite{Eling:2005zq} and also depend on the $c_n$.

Constraints on the $c_n$ are then imposed by the requirements that
$\alpha_1$ and $\alpha_2$ vanish, and that wave modes have
positive energy densities and real speeds.  Another constraint
arises from imposing superluminal speeds, which is required to a
first approximation by the absence of vacuum \v{C}erenkov
radiation of gravity-aether shock waves~\cite{Elliott:2005va}. Yet
another constraint can be derived from primordial
nucleosynthesis~\cite{Carroll:2004ai}. It was shown
in~\cite{Foster:2005dk} that all of these constraints are
satisfied by a large two parameter family of ae-theories.

Additional tests of ae-theory can come from the study of stellar
solutions and black holes, begun
in~\cite{Eling:2006df,Eling:2006ec}, and from the ``ultimate"
test~\cite{Will:1993ns} provided by observations of binary pulsar
systems. This work begins that examination with a calculation of
the generation
 of gravity-aether radiation by a nearly Newtonian source and
the subsequent energy loss, or radiation damping, of the source. A
formula is derived for the rate of change of energy:
\beq\label{RATE1}
        \frac{d \mathcal{E}}{dt} = -G_N\bigg<
    \frac{\mathcal{A}}{5}
            \Big( \frac{d^3 Q}{dt^3}\Big)^2
        + \mathcal{B} \Big(\frac{d^3 I}{dt^3}\Big)^2
        + \mathcal{C}\Big(\frac{d\Sigma}{dt}\Big)^2\bigg>,
\eeq
where $Q_{ij}$ is the trace-free quadrupole moment of the source,
$I$ is the trace of the second moment, $\Sigma_i$ is a dipolar
quantity defined below, and $\mathcal{A},\mathcal{B},$ and
$\mathcal{C}$ are dimensionless combinations of the $c_n$; $G_N$
is the value of Newton's constant that one would measure far from
an external gravitating source; the angular brackets indicate a
time average over a period of the system's motion. This formula
generalizes the ``quadrupole" formula of standard general
relativity, which predicts a similar expression but with
$\mathcal{A} = 1$, $\mathcal{B} = \mathcal{C} = 0$.

In the case of a system of two compact bodies, this expression
takes the form
\beq\label{RATE2}
     \frac{d\mathcal{E}}{dt}= -G_N
    \Big<\Big(\frac{G_N \mu m}{r^2}\Big)^2
        \Big(\frac{8}{15}\mathcal{A}
        \big(12 v^2 - 11(\frac{dr}{dt})^2\big)
        +4 \mathcal{B}\big(\frac{dr}{dt}\big)^2
        +\mathcal{C}'\mathcal{D}^2\Big)\Big>,
\eeq
where $\mu$ is the reduced mass of the system, $m$ the total mass,
$v$ the relative velocity of the bodies, and $r$ their orbital
separation, which I assume is much larger than the size $d$ of the
bodies; $\mathcal{D}$ is the difference in self-gravitational
binding energy per unit mass of the bodies, and the coefficient
$\mathcal{C}'$ is another dimensionless combination of the $c_n$.

This expression gives the lowest-order effects in a post-Newtonian
(PN) (weak field, slow motion) expansion. Aside from the $(G_N \mu
m/r^2)^2$ prefactor, the first two terms are $O(G_N m/r)$ and the
last is $O((G_N m/d)^2)$. It does not take into account strong
field effects that may be important when the fields are not weak
inside a given body. The strength of the field of a compact body
can be characterized by the quantity $(G_N m/d)$; this is ``small"
for the sun ($\sim 10^{-6}$) or a typical white dwarf ($\sim
10^{-3}$), and ``large" for a typical neutron star ($\sim
10^{-1}$) or black hole ($\sim 1$). Thus, the field should be
strong within the systems actually used to measure the damping
rate. Strong field effects on the damping rate of a compact body
can be associated with a dependence of the body's gravitating mass
on the ambient non-metric fields---that is, with a violation of
the strong equivalence principle~\cite{Will:2001mx}. These effects
are not present in GR at lower post-Newtonian orders. Their
presence in ae-theory will be examined in future
work~\cite{Foster:2007gr}.

The damping rate can be tested by observing the rate of change of
the orbital period $P$ of various binary systems, since $(d
P/dt)/P = -(3/2) (d\mathcal{E}/dt)/\mathcal{E}$, equating the
energy radiated to minus the change in mechanical energy of the
system. In practice, this test is conjoined with tests of other
``post-Keplerian" (PK)
parameters~\cite{Will:2001mx,Stairs:2003eg}, in particular the
rate of advance of periastron (the point at which the two objects
are closest to each other) and the redshift or time delay due to
the gravitational field of the system.  These
``quasi-static"~\cite{Stairs:2003eg} parameters are determined by
the post-Newtonian forms of the fields and the effective equations
of motion for the compact bodies.  The conjoint technique is
necessary, because the expressions for the PK parameters depend on
the unknown masses of the systems' bodies. The expressions for the
parameters will depend on the two masses, other measurable
parameters, and a given theory's free parameters. Measurement of
three mass dependent parameters, for fixed values of the theory
parameters, gives three bands with widths due to errors in the
two-dimensional space of mass values. The theory is consistent for
those values of the free parameters if the bands overlap. The
predictions of GR have been validated in this way using data from
various binary systems containing pulsars, whose regular pulsing
provides an accurate measuring device; see the
review~\cite{Stairs:2003eg} for details.

For ae-theory, I find that if one assumes the strong field effects
are negligible so that the results below are adequate, then there
exists a one-parameter family of theories that satisfy all of the
constraints summarized in~\cite{Foster:2005dk}, and whose
predictions for the PK parameters match those of GR to the order
worked to here. This can be seen as follows. To lowest PN order
and neglecting strong field effects, the quasi-static parameters
can be determined within the PPN framework~\cite{Will:1993ns}.
Consequently, when $\alpha_1$ and $\alpha_2$ are set to zero, so
that all of the ae-theory PPN parameters match those of GR, the
two theories will make the same predictions for the quasi-static
parameters. In this case, the $c_n$ can be constrained by
requiring that the damping rate equal that of GR. But as shown
below, the radiated fields contain only quadrupolar contributions
when $\alpha_1$ and $\alpha_2$ are set to zero. The damping rates
then coincide when $\mathcal{A}$ is set to one, which can be done
by imposing one condition on the two remaining free $c_n$.  To be
consistent with the observational tests summarized
in~\cite{Foster:2005dk}, this curve of theories must intersect
with the allowed two-parameter family demarcated there (and below
in Sec.~\ref{FIVE}).  This is the case all along the curve, as
long as $c_-,c_+ \geq 0$.

The calculation will proceed as follows. In Sec.~\ref{TWO}, a weak
field expansion of the field equations is performed. The
perturbations are shown to satisfy the wave equation, with matter
terms and nonlinear terms acting as sources.  In Sec.~\ref{THREE},
these equations are solved via integration of the sources with
Green's functions. The source integrals are approximated in terms
of time derivatives of moments of the sources, and evaluated to
order of interest using the PPN expansion of the fields.  In
Sec.~\ref{FOUR}, an expression for the rate of change of energy
contained within a volume of space is defined, and evaluated in
terms of the wave forms. I will conclude with a discussion of the
constraints on the $c_n$ implied by observations from binary
pulsars, in Sec.~\ref{FIVE}.

I use the conventions of~\cite{Wald:1984rg}.  In particular, the
metric signature is ${(-,+,+,+)}$.  This differs from the
published version of this article~\cite{Foster:2006az}; however,
this signature is vastly more convenient for performing
calculations involving a time-space decomposition, and should have
been used from the start. The coefficients in the action below
will be defined so that the results for the PPN parameters and the
damping rate match the published results. Spatial indices will be
indicated by lowercase Latin letters from the middle of the
alphabet: $i,j,k,\dots$. One exception is that the coefficients
$c_{1,2,3,4}$ will often be referred to collectively as $c_n$ in
the text, when no confusion should arise. Upon performing the
weak-field expansion of the field equations (Eqn.~\eqref{WEAK}
onwards), indices will be raised and lowered with the flat metric
$\eta_{ab}$. Repeated spatial indices will be summed over,
regardless of vertical position: $T_{ii} = \sum_{i = 1\dots 3}
T_{ii}$. The flat-space Laplacian will be denoted by $\del$: $\del
f \equiv f_{,ii}$. Time indices will be indicated by a $0$; time
derivatives will be denoted by an overdot:
$\dot{f}\equiv\partial_0 f$. I adopt units in which the flat-space
speed-of-light $c = 1$.
%
%
%
\section{Field equations}
\label{TWO}
In this section, I will expand the ae-theory field equations about
a flat metric, constant aether  background, obtaining a set of
wave equations with matter terms and nonlinear terms as sources.
\subsection{Exact equations}
I begin with the standard four-parameter ae-theory action $S$
\beq\label{ACT5}
    S = \frac{1}{16\pi G}\int d^4x\,\sqrt{|g|}\,\Big(
        R - K^{ab}_{\ph{ab}cd}\nabla_a u^c \nabla_b u^d
        +\lambda(u^a u^b g_{ab} + 1)\Big),
\eeq
where
\beq
        K^{ab}_{\phantom{ab}cd} = \big(c_1 g^{ab}g_{cd}
        +c_2\delta^a_c\delta^b_d + c_3\delta^a_d\delta^b_c
        -c_4 u^a u^b g_{cd}\big).
\eeq
The sign of the $c_4$ term looks awkward, but it allows for easy
comparison between results in this version of the article and in
the published version~\cite{Foster:2006az}.  I will use shorthand
for certain combinations of the $c_n$:
\beq
    c_{14} = c_1 - c_4,\quad c_{123} = c_1 + c_2 + c_3,\quad
        c_{\pm} = c_1 \pm c_3.
\eeq
In addition to the ae-theory action, there is an
aether-independent matter action. The matter can be assumed to
couple universally to some metric since Lorentz-violating effects
in nongravitational interactions are already highly
constrained~\cite{Mattingly:2005re,Bluhm:2005uj}. Aether couplings
are then excluded from the matter action, and the field $g_{ab}$
is identified as this universal metric.

The resulting equations of motion consist of the Einstein
equations
\beq\label{EEQ5}
    G_{ab} - S_{ab} = 8\pi G T_{ab},
\eeq
where
\beq
    G_{ab} = R_{ab} -\hlf R g_{ab},
\eeq
\beq
\begin{split}
    S_{ab} = &\nabla_c\bigl(
       K^c_{\ph{c}(a} u_{b)} + K_{(ab)}u^c - K_{(a}^{\ph{(a}c}u_{b)}\bigr)\\&
        +c_1\bigl(\nabla_a u_c\nabla_b u^c - \nabla_c u_a\nabla^c u_b
             \bigr)
        +c_4(u^c\nabla_c u_a)( u^d\nabla_d u_b)\\
        &+\lambda u_a u_b - \frac{1}{2}g_{ab}
            (K^c_{\ph{a}d}\nabla_c u^d),
\end{split}
\eeq
 with
\beq
    K^a_{\phantom{a}c} = K^{ab}_{\phantom{ab}cd}\nabla_b u^d,
\eeq
and $T_{ab}$ is the matter stress tensor. There are also the
aether field equations,
\beq\label{AEQ5}
    \nabla_b K^b_{\ph{a}a} = -\lambda u_a
        - c_4 (u^c\nabla_c u_b)\nabla_a u^b,
\eeq
and the constraint
\beq\label{CON5}
    g_{ab}u^a u^b = -1.
\eeq
Eqn.~\eqref{AEQ5} can be used to eliminate $\lambda$, giving
\beq
    \lambda = u^c\nabla_a K^a_{\phantom{a}c}
    + c_4(u^c\nabla_c u^a)(u^d\nabla_d u_a).
\eeq
\subsection{Linear-order variables}
I will now expand the exact equations about a flat background. I
assume a Minkowskian coordinate system and basis with respect to
which, at zeroth order, the metric is the Minkowski metric
$\eta_{ab}$ and the aether is purely timelike.  I then define
variables $h_{ab}$ and $w^a$, with
\beq\label{WEAK}
    h_{ab} = g_{ab} - \eta_{ab},\qquad
    w^0 = u^0 - 1,\qquad
    w^i = u^i.
\eeq
I assume that $h_{ab}$ and $w^a$ fall off at spatial infinity like
$1/r$.

I will further define variables by decomposing the above into
irreducible transverse, or ``divergence-free",
 and longitudinal, or ``curl-free", pieces.
The decomposition is unique and well-defined in Euclidean space,
having imposed the above boundary conditions (one is essentially
solving Laplace's equation---see~\cite{Arnowitt:1962hi} for more
discussion). First, consider the spatial  vectors $w^i$ and
$h_{0i}$, and define the following variables:
\beq
    h_{0i} = \gamma_i + \gamma_{,i}\quad
    w^i = \nu^i + \nu^{,i},
\eeq
with $\gamma_{i,i} = \nu^{i}_{,i} = 0$. Next, consider the spatial
components of the metric $h_{ij}$.  A symmetric, 2-index tensor on
Euclidean space can be uniquely decomposed into a
transverse-trace-free tensor, a transverse vector, and two scalar
quantities representing the transverse and longitudinal traces:
\beq
    h_{ij} = \phi_{ij} + \hlf P_{ij}[f] + 2\phi_{(i,j)} + \phi_{,ij},
\eeq
where
\beq
    0=\phi_{ij,j} = \phi_{jj} = \phi_{i,i},
\eeq
and
\beq
    P_{ij}[f] = \delta_{ij}\del f - f_{,ij};
\eeq
hence, $P_{ij}[f]_{,j} = 0$, and $h_{ii} = \del( f + \phi)$.
Further define
\beq
    F = \del f.
\eeq
The list of variables then consists of  a transverse-traceless
spin-2 tensor $\phi_{ij}$, transverse spin-1 vectors
$\gamma_i,\nu^i,\phi_i$, and spin-0 scalars
$h_{00},w^0,\gamma,\nu,F,\phi$.

I will impose coordinate gauge conditions below, after expressing
the field equations in unfixed form. The standard gauge to impose
when performing the analogous calculation in conventional GR is
the ``harmonic" or ``Lorentz" gauge, $2h_{ab}^{\ph{ab},b} =
\eta^{cd}h_{cd,a}$, as this happens to reduce the field equations
to a simple form when they are expressed covariantly. Some variant
of this condition has a similar effect in several other
alternative theories of gravity, as seen in~\cite{Will:1977wq}.
Here, the increased complexity of the equations and the
noncovariant decomposition of them and the field variables means
that no obvious extension of the harmonic gauge has such a
utility. Instead, the gauge will be chosen somewhat arbitrarily so
as to eliminate certain variables:
\beq
    0 = w_{i,i} = h_{0i,i} = h_{i[j,k]i},
\eeq
or equivalently,
\beq\label{GGE5}
    0 = \nu = \gamma = \phi_i.
\eeq

An infinitesimal coordinate gauge transformation has the
linear-order form
\beq
    \delta h_{ab} = \xi_{a,b} + \xi_{b,a}\quad
    \delta w^a = -\dot{\xi}^a.
\eeq The conditions~\eqref{GGE5} can be realized while in an
arbitrary gauge (a prime denotes that the variables are evaluated
in the original gauge) by choosing $\xi_0 = -(\gamma ' + \nu ')$
and the transverse part of $\xi_i$ as $-\phi'_i$, and by solving
for the longitudinal part $\xi$ of $\xi_i$ via $\dot{\xi} = \nu'$.
One constraint on the choice of gauge is that it must be a valid
PPN gauge, as defined in~\cite{Will:1993ns}, so that the integrals
of Sec.~\ref{SIGMA} can be evaluated by expressing the variables
in terms of their PPN expansion. The above is a valid, albeit
nonstandard, PPN gauge (in contrast to the gauge chosen
in~\cite{Jacobson:2004ts,Eling:2005zq}).
\subsection{Linearized equations}
I now express the field equations~\eqref{EEQ5} and~\eqref{AEQ5} in
terms of the above variables, and arrange them in the form
\begin{gather}
    \bar{G}_{ab} - \bar{S}_{ab} = 8\pi G\big(T_{ab} + t_{ab}\big),\\
    \label{AELIN}\bar{K}^b_{\ph{a}a,b} = 8\pi G\sigma_a,
\end{gather}
where the overbar denotes the portion of the tensor linear in
$h_{ab}$ and $w^a$, and the nonlinear source terms $t_{ab}$ and
$\sigma_a$ are defined to a given order in the variables by
asserting that the above equations equal the exact equations to
that order. It will prove convenient to combine the equations in
the form
\beq\label{RELAX}
    \bar{G}_{ab} -\bar{S}_{ab}
    + \delta^0_{[a}\bar{K}^c_{\ph{a}b],c}
        = 8 \pi G \tau_{ab},
\eeq
thus defining the source
\beq\label{DEFSOURCE}
    \tau_{ab} = T_{ab} + t_{ab} + \delta^0_{[a}\sigma_{b]}.
\eeq
Identities satisfied by the linear-order terms will imply
conservation of $\tau_{ab}$.

The constraint~\eqref{CON5} to linear order is
\beq\label{CONLIN}
    w^0 = \hlf h_{00}.
\eeq
I will use this result to eliminate $w^0$. The form of nonlinear
terms will not be needed as explained in Sec.~\ref{SIGMA}.

Now,
\beq
       \bar{G}_{ab} = -\hlf (\del h_{ab}- \ddot{h}_{ab}) - \hlf h_{,ab}
                + h_{c (a,b)}^{\ph{c(a,b)}c}
                +\hlf\eta_{ab}(\del h- \ddot{h}
           -h_{cd}^{\ph{cd},cd}),
\eeq
where $h= \eta^{ab}h_{ab}$.  Hence,
\beq
     \bar{G}_{ij} = -\hlf\big[ \del\phi_{ij}-\ddot{\phi}_{ij}  \big]
                +\big[\ddot{\phi}_{(i,j)} - \dot{\gamma}_{(i,j)}\big]
        +\frac{1}{4}P_{ij}[\del f-\ddot{f}  - 2h_{00} - 2\ddot{\phi}
        +4\dot\gamma]
        -\hlf\ddot{f}_{,ij},
\eeq
\beq
        \bar{G}_{0i} = -\hlf\del(\gamma_i - \dot{\phi}_i) -
        \hlf(\dot{F})_{,i},
\eeq
\beq
        \bar{G}_{00} = -\hlf \del F.
\eeq

The linear-order forms of the covariant derivatives of $u_a$ are
\beq
\begin{split}
       \over{\nabla_0 u^i} &= \dot{w}^i+\dot{h}_{0i}-\hlf h_{00,i}\\
                &= \dot{\nu}^i+\dot{\gamma}_i
            +(\dot\nu+\dot\gamma-\hlf h_{00})_{,i},
\end{split}
\eeq
\beq
\begin{split}
        \over{\nabla_i u^j} &= w^j_{,i}+ h_{0[j,i]} + \hlf \dot{h}_{ij}\\
                &= \hlf \dot{\phi}_{ij}
            + \nu^j_{,i} + \gamma_{[j,i]} + \dot{\phi}_{(j,i)}
            + \frac{1}{4}P_{ij}[\dot{f}]
            +(\nu + \hlf \dot{\phi})_{,ij},
 \end{split}
\eeq and $\over{\nabla_a u^0} = 0$.

From
\beq
        \bar{S}_{ab} = \dot{\bar{K}}_{(ab)}
     + \delta^0_{(a}\bar{K}_{b)\ph{c},c}^{\ph{b)}c},
\eeq
follows
\beq
\begin{split}
        \bar{S}_{ij} &= \partial_0
        (c_+\over{\nabla^{(i}u^{j)}}+c_2\delta_{ij}\over{
    \nabla_k u^k})\\
               & = \frac{c_+}{2} \ddot{\phi}_{ij} +
        c_+(\dot\nu^{(i,j)}
                +\ddot{\phi}_{(i,j)})
        +\hlf P_{ij}[c_2(2\dot\nu+\ddot{\phi} + \ddot{f})
        +\frac{c_+}{2}\ddot{f}]\\
       &\quad +\hlf\big((c_2+c_+)(2\dot\nu+\ddot{\phi})
            +c_2\ddot{f}\big)_{,ij},
\end{split}
\eeq
\beq\begin{split}
        \bar{S}_{0i} + \hlf\bar{K}_{ai}^{\ph{ai},a} &= \bar{K}_{(ij),j}
    = c_{+}\partial_j(\over{\nabla^{(i}u^{j)}})\\
        &=
        \hlf\del\Big(c_+(\nu^i + \dot{\phi}_i)
                +\big((c_+ + c_2)(2\nu+\dot{\phi}) + c_2
        \dot{f}\big)_{,i}\Big),
\end{split}
\eeq
\beq\begin{split}
        \bar{S}_{0i} - \hlf \bar{K}_{ai}^{\ph{ai},a}
        &= \dot{\bar{K}}_{0i} + \bar{K}_{[ij],j}
        =c_{14}\partial_0(\over{\nabla_0 u^i})
        +c_- \partial_j(\over{\nabla^{[i}u^{j]}})\\
            &= c_{14}(\ddot{\nu}^i + \ddot{\gamma}_i)
        - \frac{c_-}{2}\del (\nu^i + \gamma_i)
        + c_{14}\big(\ddot{\nu}+\ddot{\gamma}
            -\hlf \dot{h}_{00}\big)_{,i}
\end{split}\eeq
and
\beq
        \bar{S}_{00} = c_{14}\partial_j(\over{\nabla_0 u^j}) =
        \del\Big( c_{14}(\dot\nu
                +\dot\gamma - \hlf h_{00})\Big).
\eeq

The above expressions indicate that the linear-order terms satisfy
the identity
\beq
    \big(\bar{G}_{ab} - \bar{S}_{ab}
        + \delta^0_{[a}\bar{K}^c_{\ph{c}b],c}\big)^{,b} = 0.
\eeq
This implies that the source $\tau_{ab}$~\eqref{DEFSOURCE} obeys a
conservation law
\beq
    \tau_{ab}^{\ph{ab},b} = \tau_{ai,i}-\dot{\tau}_{a0}  = 0.
\eeq
\subsection{Wave equations}
The above equations can now be decomposed as the variables. The
field equations for variables of different spin will separate. I
will impose below the gauge conditions~\eqref{GGE5}. The following
results are equivalent to those of~\cite{Jacobson:2004ts}
expressed in a different gauge when $\tau_{ab}=0$; in particular,
the count of independent plane wave modes---two spin-2, two
spin-1, and one spin-0---and the expressions for the wave speeds
are recovered.

I now consider the different spins in turn.  Define the set of
operators
\beq
    \Box_i \psi \equiv \del\psi -(s_i)^{-2} \ddot{\psi} ,
\eeq
for $i = 0,1,2$.
\subsubsection{Spin-2}
The transverse-traceless part of the space-space components
of~\eqref{RELAX} gives
\beq\label{BOX2}
        {\Box}_2 \phi_{ij} = -16\pi G \tau_{ij}^{\textrm{TT}},
\eeq
with
\beq
        (s_2)^2 = \frac{1}{1-c_+},
\eeq
and where TT signifies the transverse-traceless projection.
\subsubsection{Spin-1}
Now the spin-1 variables. The transverse parts of~\eqref{RELAX}
give
\beq\label{DEL1}
        \del( c_+\nu^i + \gamma_i) = -16\pi G\tau_{i0}^{\textrm{T}}, \eeq
and
\beq
        c_{14}(\ddot{\nu}^i+\ddot{\gamma}_i)-\hlf\del(c_-\nu^i
        + (1-c_-)\gamma_i)= -8\pi G \tau_{0i}^{\textrm{T}}.
\eeq
where the T signifies the transverse projection. These relations
imply
\beq\label{BOX1}
        {\Box}_1 (\nu^i + \gamma_i)  = \frac{-16\pi G}{2c_1 - c_+c_-}
            \big(c_+ \tau_{i0} +(1-c_+) \sigma^i\big)^{\textrm{T}},
\eeq
with
\beq
        (s_1)^2 = \frac{2c_1 - c_+c_-}{2(1-c_+)c_{14}}.
\eeq
\subsubsection{Spin-0}
Now consider the spin-0 variables. The transverse-trace and
longitudinal-trace portions of the space-space components
of~\eqref{RELAX} give
\beq
    (1+2c_2+c_+)\ddot{F} - \del \big(F -2 h_{00} -
    2(1+c_2)\ddot{\phi}\big)
        = -16\pi G\tau^{\textrm{T}}_{ii},
\eeq
and
\beq
        (1+c_2)\ddot{F} + c_{123}\del\ddot{\phi}
         = -16\pi G\tau^{\textrm{L}}_{ii},
\eeq
where $\tau_{ij}^{\textrm{L}} = \tau_{ij} -
\tau^{\textrm{T}}_{ij}$. The time-time component of~\eqref{RELAX}
gives
\beq\label{DEL2}
        \del(F - c_{14}h_{00}) = -16\pi G\tau_{00},
\eeq
and the longitudinal space-time component of~\eqref{RELAX} gives
\beq\label{DEL3}
    \del\big((1+c_2)\dot{f} + c_{123}\dot{\phi}\big)_{,i}
        = -16\pi G\tau^{\textrm{L}}_{i0},
\eeq
where $\tau^{\textrm{L}}_{i0} = \tau_{i0} -
\tau^{\textrm{T}}_{i0}$. These equations imply
\beq\label{BOX0}
    \Box_0 F = -\frac{16\pi G c_{14}}{2-c_{14}}(\tau_{ii}
        -\frac{2+3c_2+c_+}{c_{123}}\tau_{ii}^{\textrm{L}}
        +\frac{2}{c_{14}}\tau_{00}),
\eeq
with
\beq\label{URG}
        (s_{0})^2 =
    \frac{(2-c_{14})c_{123}}{(2+3c_2+c_+)(1 - c_+)c_{14}}.
\eeq
Further implied by these and the untraced, transverse-trace part
of~\eqref{RELAX} is the equation
\beq
    \Box_0 f_{,ij} = \tau_{ij}'.
\eeq
The form of the source $\tau'_{ij}$ is unimportant; only the fact
that $f_{,ij}$ satisfies a sourced wave equation is needed so that
later eqn.~\eqref{ZONES} can be applied when evaluating the
damping rate expression in Sec.~\ref{FOUR}.
\section{Evaluation of source integrals}
\label{THREE}
The above equations can be formally solved via integration of the
sources with the appropriate Green's function, and the resulting
integrals approximated in terms of time derivatives of moments of
the source.  Upon doing so, the nonstatic contributions to the
fields to desired accuracy at points far from the material source
depend on two integral quantities, the second mass moment of the
material source
\beq
    I_{ij} = \int d^3 x \rho\, x_i x_j,
\eeq
where $\rho = T_{00}$ to lowest order, and the integral
\beq
    \Sigma_i = \int d^3 x\, \sigma_i,
\eeq
where $\sigma_i$ are the quadratic terms from the aether field
equation~\eqref{AELIN}.
\subsection{Approximation of source integrals}
Equations of the form
\beq\label{GENERIC}
   \del \psi -(s)^{-2} \ddot{\psi}   = - 16\pi \tau,
\eeq
can be solved with outward-going disturbances at infinity by
writing
\beq\label{SOLN}
    \psi(t,\mathbf{x}) = 4\int d^3 x'
    \frac{\tau(t-z/s,\mathbf{x}')}{z},
\eeq
where $z = |\mathbf{x}-\mathbf{x}'|$.

The source integral can be simplified with a standard
approximation~\cite{Will:1993ns}. As indicated by the energy loss
rate expression~\eqref{EDOT}, only the portion of the fields that
fall off as $(1/r)$ are of interest.  A weak field, slow motion
assumption will be made: the material source should be described
by a mass $m$, a size $L$, and a time-scale $T$ such that $(G_N
m/L)$ and $ (L/T s)$ are small quantities. Then only terms of
interesting order are retained in the following expansion:
\beq\label{APPROX}
    \psi(t,\mathbf{x})\approx \frac{4}{R}
    \Big(\sum^\infty_{m=0} \frac{1}{m!s^m}
            \frac{\partial^m}{\partial t^m}
            \int \tau(t-R/s,\mathbf{x}')\big(x'^i\hat{x}^i\big)^m
    \Big),
\eeq
where $R = |\mathbf{x}|$ and $\hat{\mathbf{x}} = \mathbf{x}/R$,
and $R \gg L$.

Now, the following sleight of hand justifies solving the
decomposed ae-theory equations by first approximating the integral
on the right side of~\eqref{SOLN} using the full $\tau$ and
\emph{then} taking the projection. Introduce the notation
$[[\tau(t,\mathbf{x}')]]\equiv \tau(t-z/s,\mathbf{x}')$. Because
the quantity on the right side of~\eqref{SOLN} depends on
$\mathbf{x}$ only through $z$, it follows that
\beq
    \frac{\partial}{\partial x^i}\int \frac{[[\tau(t,\mathbf{x}')]]}{z} =
            -\int\frac{\partial}{\partial x'^i}
        \big(\frac{[[\tau(t,\mathbf{x}')]]}{z}\big) +
                \int\frac{[[{\partial_i}'\tau(t,\mathbf{x}')]]}{z}.
\eeq
It follows from this that, e.g.,
\beq
    \int\frac{[[\tau_{ij}^{\textrm{T}}(t,\mathbf{x}')]]}{z} =
        \bigg(\int\frac{[[\tau_{ij}(t,\mathbf{x}')]]}{z}\bigg)^{\textrm{T}},
\eeq
after discarding integrals of total derivatives, where T on the
left side signifies transverse with respect to $\mathbf{x}'$, and
on the right side transverse with respect to $\mathbf{x}$.  As a
further convenience, it follows that  to $O(1/R)$, the transverse
projection is equal to the algebraic projection in the direction
orthogonal to $\hat{\mathbf{x}}$.

Additionally, there are the Poissonnian
equations~\eqref{DEL1},~\eqref{DEL2}, and~\eqref{DEL3} of the form
\beq
    \del \psi = -16\pi\tau.
\eeq
Solving via Green's function and expressing to $O(1/R)$ far from
the source gives
\beq
    \psi(t,\mathbf{x}) \approx \frac{4}{R} \int d^3x'
    \tau(t,\mathbf{x}').
\eeq
The integrals of the sources in these particular equations happen
to be conserved quantities. Thus, ignoring static terms in the
wave forms, the equations are effectively \textit{un}sourced, and
thus imply
\beq
    \psi = 0,
\eeq
to sufficient accuracy.
\subsection{Sorting of source integrals}
To evaluate the integrals indicated by Eqn.~\eqref{APPROX}, I will
express the sources in terms of their post-Newtonian expansions.
I will further assume that the system is composed of compact
bodies of individual size $d \ll L$ that exert negligible tidal
forces on each other.   The system will then have an orbital
velocity $v\sim \sqrt{G_N m/L}$. Following the discussion
in~\cite{Will:1993ns}, the leading-order terms in the fields will
be $O(G_N m^2/L)$, which give the quadrupolar and monopolar
contributions, and $O((G_Nm^2/d)v)$, giving the dipolar
contribution. Terms of these orders can only result from integrals
of terms that are, respectively, 2PN and 2.5PN order. Integrals of
interest can be identified by noting that since the rate of change
of the system is governed by its velocity, assumed to be .5PN
order, taking the time derivative of a quantity effectively
multiplies it by a factor of $v$ and raises it by .5PN orders.
Also, only nonstatic, or non-conserved, terms are of interest as
only the time derivatives of the fields will appear in the
expression for the energy loss.

I begin by considering the moments of $\tau_{ij}$.  First, the
conservation law implies
\beq
    \int\tau_{ij} = \hlf\int\ddot{\tau}_{00} {x'}_i {x'}_j
    + \int \dot{\sigma}_{(i} {x'}_{j)}
    =\hlf\int\ddot{T}_{00} {x'}_i {x'}_j
    =\hlf \ddot{I}_{ij},
\eeq
where the last two equalities hold to desired order and for the
last I have used the Eulerian continuity equation for the fluid
\beq\label{CONT}
    \dot{\rho} + (\rho v^i)_{,i} = 0,
\eeq
assumed to hold at $O(1.5)$. Then,
\beq \int\dot{\tau}_{ij}x'_k = -\hlf\int\big(\ddot{\tau}_{i0}
                                        x_j x_k
    + \ddot{\tau}_{j0}x_k x_i
        -\ddot{\tau}_{k0}x_i x_j\big),
\eeq
which is of uninteresting order, as are remaining moments.

I now consider the moments of $\tau_{i0}$.  The integral  of
$\tau_{i0}$ is conserved, so I ignore it.  Next,
\beq
    \int\dot{\tau}_{i0}x_j = -\int \tau_{ij}
        = -\hlf\ddot{I}_{ij}.
\eeq
The other moments are of uninteresting orders.

I now consider moments of $\tau_{00}$.  First, the integral of
$\tau_{00}$ is conserved, so I ignore it.  Then,
\beq
    \int\dot{\tau}_{00} x_i = -\int\tau_{0i}
            = -\Sigma_i,
\eeq
where the second equality ignores the static integral of
$\tau_{i0}$. Finally,
\beq
    \int\ddot{\tau}_{00}x_i x_j = \ddot{I}_{ij},
\eeq
to desired order.
\subsection{Evaluating $\Sigma_i$}
\label{SIGMA}
I now consider the moments of $\sigma_i$. The terms in $\sigma_i$
are at least 2.5PN order. The only integral of interest is thus
$\Sigma_i = \int \sigma_i$, which is $O((G_N m^2/d)v)$.  At this
point,  I can explain why the nonlinear terms in the unit
constraint~\eqref{CON5} can be ignored.  The previous subsection
makes clear that only their appearance in $\sigma_i$ need be
considered. As follows from the PN forms given in Chapter 4, the
nonlinear constraint terms are integer PN orders starting with 2PN
and have no free indices, and there are no field variables that
are .5PN order. It follows that any nonlinear constraint terms
appearing at 2.5PN order in $\sigma_i$ must do so in the form
$(terms)_{,0i}$. Total derivatives do not contribute to
$\Sigma_i$, so these terms can be ignored.

I will evaluate $\Sigma_i$ explicitly by expressing the fields in
terms of the PPN expansion, but in the nonstandard coordinate
gauge~\eqref{GGE5}. The PPN forms in the standard PPN coordinate
gauge, adjusted to the conventions used here, are reported
in~\cite{Foster:2005dk} as
\begin{gather}
    \phi_{ij} = 0\\
    \gamma_i = -\frac{2c_1}{c_-}\nu^i
        = -\frac{8+\alpha_1}{4}(V_i + W_i),
        \qquad \phi_i = 0\\
    h_{00} = -\del\chi,\quad f=2\phi=-2\chi,\\
        \gamma = -\frac{1}{4}(6+ \alpha_1 - 2\alpha_2)\dot\chi,
        \quad\nu = \frac{2c_1 + 3c_2 + c_3 + c_4}{2
        c_{123}}\dot\chi,
\end{gather}
where
\begin{gather}
    V_i(x) = G_N\int d^3x' \frac{\rho(x')v^i}{z},
    \qquad W_i(x) = G_N\int d^3x' \frac{\rho(x) v_j z_j z^i}
            {z^3},\\
    \chi(x) = -G_{N}\int d^3 x' \,\rho(x')\,z,
\end{gather}
with $z^i = x^i - x'^i$ (note that $\dot\chi = (V_i - W_i)_{,i}$),
and
\begin{gather}
    G_N = \frac{G}{1-(c_{14}/2)},\\
    \alpha_1 = -\frac{8(c_1 c_4 + c_3^2)}{2c_1 - c_+ c_-},\label{alph1}\\
    \alpha_2 = \frac{\alpha_1}{2} - \frac{(c_1 + 2c_3 - c_4)
                (2c_1+3c_2+c_3+c_4)}{(2-c_{14})c_{123}}.\label{alph2}
\end{gather}
Adjustment from the standard to the nonstandard gauge is done by
defining a gauge parameter $\xi_a$ with $\xi_0 = -(\gamma'+
\nu')$, $\dot{\xi}_i = \nu'_{,i}$, where $\gamma', \nu'$ are the
standard-gauge values. Then in the nonstandard gauge, the
variables are as above except $\nu=\gamma=0$ and
\beq
    \phi = \frac{(c_1 + 2c_2 + c_4)}{c_{123}}\chi.
\eeq

With these forms, $\Sigma_i$ can be evaluated, and after some
algebra gives
\beq
    \Sigma_i = \frac{1}{2}\int \rho
        \big((\alpha_1-\alpha_2)V_i + \alpha_2 W_i\big).
\eeq
I will later consider the special cases of a single, compact,
spherically symmetric body and of a pair of compact bodies that
are static and spherically symmetric in their own rest frames. In
the first case, spherical symmetry implies that $\Sigma_i$
vanishes---otherwise it would define a symmetry-breaking spatial
vector. In the second case, or more generally with $n$ such
bodies, $\Sigma_i$ can be simplified via the following
observations. First, the Newtonian potential $U$ felt at a given
body contains an $O(G_N m/L)$ contribution from the presence of
the other bodies, plus an $O(G_N m/d)$ self-contribution
$\bar{U}$,
\beq
    \bar{U}_a(x_a) = G_N\int_a d^3x'\frac{\rho}
            {|\mathbf{x_a}-\mathbf{x}'|},
\eeq
where the integral extends just over the ``$a$-th" body.  Second,
spherical symmetry of each body implies that $(\Omega_a)_{ij} =
(1/3)\Omega_a\delta_{ij}$, where
\beq
    (\Omega_a)_{ij}\equiv
    -\hlf G_N\int_a d^3 x\,d^3 x' \frac{\rho(x)\rho(x')z_i z_j}{z^3},
\eeq
and
\beq
    \Omega_a = (\Omega_a)_{ii}
        = -\hlf\int_a d^3x \rho \bar{U}_a.
\eeq
Third, staticity of each body implies that $V_i = \sum_a (v_a)^i
U$, and similarly for $W_i$.  These facts imply that
\beq
    \int \rho V_i = 3\int\rho W_i = -2 \sum_a (v_a)^i \Omega_a,
\eeq
plus terms of $O(G_N m^2 v/L)$. Therefore, to interesting order,
\beq\label{MMM}
    \Sigma_i =-(\alpha_1 - \frac{2}{3}\alpha_2)
    \sum_a (v_a)^i \Omega_a.
\eeq
\subsection{Wave forms}
\label{WAVEFORM}
I can now express the nonstatic, radiation-zone fields, to desired
accuracy.  For spin-2,
\beq
    \phi_{ij} = \frac{2G}{R}\big(\ddot{Q}_{ij}\big)^{\textrm
    {TT}}.
\eeq
where
\beq
    Q_{ij} = I_{ij} - \frac{1}{3}I,\qquad I = I_{ii}.
\eeq
For spin-1,
\begin{gather}
    \nu^i = \frac{-2G}{R}\frac{1}{2c_1 - c_+c_-}
        \Big(\frac{c_+}{(1 - c_+)s_1}\ddot{Q}_{ij}\hat{x}_j
        - 2 \Sigma_i\Big)^{\textrm T},\\
        \gamma_i = -c_+\nu^i,
\end{gather}
For spin-0,
\begin{gather}
    F = \frac{2G}{R}\frac{c_{14}}{2 - c_{14}}\Big(
        3(Z-1) \hat{x}^i\ddot{Q}_{ij}\hat{x}^j
    +Z\ddot{I}
        + \frac{4}{c_{14}s_0}\Sigma_i\hat{x}^i\Big),\\
    h_{00} = \frac{1}{c_{14}}F,\\
    \dot{\phi}_{,i} = -\frac{(1+c_2)}{c_{123}} \dot{f}_{,i},
\end{gather}
where
\beq\label{ZZZ}
    Z=\frac{2(\alpha_1-2\alpha_2)(1-c_+)}{3(2c_+ - c_{14})}.
\eeq
\section{Energy loss formulas}
\label{FOUR}
I now turn to the expression for the rate of change of energy
contained within a volume of space. Such an expression can be
derived via the Noether charge method for defining the total
energy of an asymptotically-flat space-time~\cite{Iyer:1994ys},
using the ae-theory Noether current derived
in~\cite{Foster:2005fr}. One can equivalently work in terms of
pseudotensors, using the results of~\cite{Eling:2005zq}.

Following the discussion in the appendix of~\cite{Iyer:1994ys}, an
expression for the total energy $\mathcal{E}$ contained in a
volume of space, for a theory linearized about a flat background,
is given by the integral over that volume of a certain
differential 3-form $\mathbf{J}_{abc} \equiv J^d
\mathbf{\epsilon}_{dabc}$. $\mathbf{J}$ can be obtained from the
quadratic order ae-theory Lagrangian modulo a boundary term. It
will depend on the metric, aether, and an arbitrary background
vector field. To define the energy, choose the background vector
field as $t^a = (\partial/\partial t)^a$. Choose the volume $V$ to
be that contained within a sphere of coordinate radius $R$.  Then,
\begin{gather}
    \mathcal{E} \equiv \int_V \mathbf{J}[t] = \int d^3 x J^0[t],\\
    \dot{\mathcal{E}}  \equiv \int \mathcal{L}_t \mathbf{J}[t]
        =\int d(t\cdot \mathbf{J}[t])
        = -\int_R d\Omega R^2 \hat{x}^i J^i[t] ,
\end{gather}
where in the second line I have used the formula $\mathcal{L}_t
\mathbf{J} = d(t\cdot \mathbf{J}) + t\cdot d\mathbf{J}$ and the
fact that $d\mathbf{J} = 0$ when the dynamical fields satisfy the
equations of motion~\cite{Iyer:1994ys}.

I will define $\mathbf{J}$ with respect to the ae-theory
Lagrangian $\mathcal{L}$ modulo a total derivative:
\beq
\begin{split}
    \mathcal{L}'\equiv&
   \mathcal{L} - \frac{1}{16 \pi G}\Big(\sqrt{|g|}
        \big(\Gamma^c_{ab} g^{ab}
            -\Gamma^b_{ab} g^{ac}\big)\Big)_{,c}\\
   & =\frac{\sqrt{|g|}}{16\pi G}
        \Big(g^{ab}\big(\Gamma^c_{ad}\Gamma^d_{cb} -
            \Gamma^c_{cd}\Gamma^d_{ab}\big)
        - K^{a}_{\ph{a}b}\nabla_a u^b\Big).
\end{split}
\eeq
The procedure of~\cite{Iyer:1994ys} gives:
\beq
\begin{split}\label{edot}
    J^a = \frac{1}{16\pi G}
        \big[&\dot{h}_{bc}\big(h^{ab,c} - \hlf h^{bc,a} + \hlf
        \eta^{bc}(h^{d,a}_d - h^{ad}_{\ph{a},d}) - \hlf
        g_d^{d,b}\eta^{ca}\\
        &- u^a K^{bc} - 2 K^{[ab]}u^c\big) - 2 \dot{u}^b
        K^{a}_{\ph{a}b}\big] - t^a \mathcal{L}',
\end{split}
\eeq
where all indices on $h_{ab}$ are raised with the flat metric
$\eta^{ab}$.

I will presume that only the time average of the damping rate need
be determined. It is then crucial to note that the damping rate is
calculated to lowest nonvanishing PN order by treating the system
as exactly Newtonian. The motion of the system can then be
decomposed into a uniform translation of the center of
mass--recall the conservation of $\int \tau_{i0}$--and a fixed
Keplerian orbit in the center-of-mass frame. As indicated below,
the field wave forms do not depend on the center-of-mass motion.
It then follows that if the time average is taken over an orbital
period, total time derivatives in~\eqref{edot} do not contribute.

It is also useful to note that approximation of the source
integrals~\eqref{APPROX} implies that to $O(1/R)$,
\beq\label{ZONES}
    \psi_{,j} = -(1/s) \dot{\psi}\hat{x}^j,
\eeq
for field $\psi$ satisfying the sourced wave
equation~\eqref{GENERIC}. This relation then implies that
\beq
    0 = \hat{x}^i \dot{\phi}_{ij}(x)
        = \hat{x}^i\dot{\nu}_i(x)
        = \hat{x}^i P_{ij}[\dot{f}(x)].
\eeq
These facts permit manipulation of terms within the integral,
e.g.:
\beq
    \int<\dot{h}_{jk} \phi_{,jki}>
    =\int<-\frac{1}{s_0}\dot{\phi}_{,jk}\dot{\phi}_{,ji}\hat{x}_k>
    =\int<-\frac{1}{s_0}\del\dot{\phi}\del\dot{\phi}\hat{x}_i>,
\eeq
where the angular brackets denote the time average.

The energy loss rate then evaluates to
\beq\label{EDOT}
    \dot{\mathcal{E}} = \frac{-1}{16\pi G}\int_R d\Omega R^2
   \Big<\frac{1}{2s_2} \dot{\phi}_{jk}\dot{\phi}_{jk}
        + \frac{(2c_1-c_+c_-)(1-c_+)}{s_1}\dot{\nu}^j\dot{\nu}^j
        + \frac{2-c_{14}}{4c_{14}s_0}\dot{F}\dot{F}\Big>.
\eeq
The sign of the coefficient of the term for each spin is opposite
to the sign of the energy density associated with linearized plane
waves, as found in~\cite{Eling:2005zq}. Thus, a positive energy
mode implies energy loss due to radiation of that mode.

The energy loss rate can be further evaluated by substituting in
the expressions for the fields given in Sec.~\ref{WAVEFORM}, and
performing the angular integral, using the results
\beq
    \frac{1}{4\pi}\int d\Omega\,
    \dot{\phi}_{ij}\dot{\phi}_{ij} = \frac{8G^2}{5R^2}
        \big(\dddot{Q}_{ij}\dddot{Q}_{ij}\big),
\eeq
\beq
        \frac{1}{4\pi}\int d\Omega
        \dot{\nu}_i\dot{\nu}_i = \frac{4G^2}{R^2}
        \frac{1}{(2c_1 - c_+c_-)^2}
        \bigg(\frac{1}{5}\Big(\frac{c_+}{(1 - c_+)s_1}\Big)^2
        \Big(\dddot{Q}_{ij}\dddot{Q}_{ij}\Big)
        +\frac{8}{3}\Big(\dot{\Sigma}_i\dot{\Sigma}_i\Big)\bigg),
\eeq
\beq
        \frac{1}{4\pi}\int d\Omega
    \dot{F}\dot{F}
    = \frac{4G^2}{R^2}\Big(\frac{c_{14}}{2 - c_{14}}\Big)^2
       \Big(\frac{6(Z-1)^2}{5}\big(\dddot{Q}_{ij}\dddot{Q}_{ij}\big)
         +Z^2\big(\dddot I \dddot I\big)
    +
    \frac{16}{3(c_{14}s_0)^2}\big(\dot{\Sigma}_i\dot{\Sigma}_i\big)\Big),
\eeq
substituted into expression~\eqref{EDOT} gives
\beq\label{EEE}
    \dot{\mathcal{E}} = -G_N\bigg<
    \frac{\mathcal{A}}{5}
            \Big(\dddot Q_{ij}\dddot Q_{ij}\Big)
        + \mathcal{B} \Big(\dddot I \dddot I\Big)
        + \mathcal{C}\Big(\dot{\Sigma}_i\dot{\Sigma}_i\Big)\bigg>,
\eeq
where
\begin{gather}\label{ABC5}
    \mathcal{A} = \Big(1 - \frac{c_{14}}{2}\Big)
    \Big(\frac{1}{s_2}
        + \frac{2c_{14}c_+^2}{(2c_1 - c_+c_-)^2}\frac{1}{s_1}
        + \frac{3(Z-1)^2 c_{14}}{2(2 - c_{14})}\frac{1}{s_0}\big),\\
\label{BBB5}
    \mathcal{B} = \frac{Z^2c_{14}}{8}\frac{1}{s_0},\\
    \mathcal{C} = \frac{2}{3c_{14}}\Big(\frac{2 - c_{14}}{s_1^3}
            +\frac{1}{s_0^3}\Big),
\end{gather}
where $Z$ is given in~\eqref{ZZZ}. This constitutes the
generalization to ae-theory of the quadrupole formula of general
relativity, and contains additional contributions from monopolar
and dipolar sources.

The presence of the monopolar term means that a spherically
symmetric source, such as a spherically pulsating star, can
radiate at this lowest nontrivial PN order in the presence of the
aether, whereas it would not in pure GR. In this case, $I_{ij} =
(1/3) \delta_{ij} I$, and as observed above, $\Sigma_i = 0$. Only
the spin-0 radiation fields are nonvanishing, and the energy loss
rate is $\dot{\mathcal{E}} = -G_N \mathcal{B}\Big<(\dddot
I)^2\Big>$. Bounds discussed in Sec.~\ref{FIVE}, however, require
that the PPN parameters $\alpha_1$~\eqref{alph1} and
$\alpha_2$~\eqref{alph2}, hence $\mathcal{B}$~\eqref{BBB5}, vanish
for observationally viable ae-theories.

For a binary system, treating the two bodies as static and
spherically symmetric in their own rest frames leads to
\beq
    \dddot{I}_{ij} = -\frac{2G_N \mu m}{r^2}
    (4 \hat{r}_{(i}v_{j)} - 3 \hat{r}_i \hat{r}_j \dot{r}),
\eeq
where $ m = m_1 + m_2$, $\mu = m_1 m_2/m$, $\mathbf{r} =
\mathbf{r}_1 - \mathbf{r}_2$ is the relative separation and
$\mathbf{v} = \dot{\mathbf{r}}$. Then, \beq
        \dddot Q_{ij}\dddot Q_{ij}
        = \frac{8}{3}(\frac{G_N \mu m }{r^2})^2(12 v^2 - 11
        \dot{r}^2).
\eeq
Also,
\beq
    \dot{\Sigma}_i = (\alpha_1 - \frac{2}{3}\alpha_2)
            \frac{G_N\mu m}{r^2}\mathcal{D} \hat{x}_i,
\eeq
where $\mathcal{D}$ is the difference in binding energy per unit
rest mass:
\beq
    \mathcal{D} = \frac{\Omega_a}{m_a} - \frac{\Omega_b}{m_b}.
\eeq
Therefore,
\beq\label{EEB}
     \dot{\mathcal{E}}= -G_N
    \Big<\Big(\frac{G_N\mu m}{r^2}\Big)^2
    \Big(\frac{8}{15}\mathcal{A}
        \big(12 v^2 - 11(\dot{r})^2\big)
        +4 \mathcal{B}\big(\dot{r}\big)^2
        +(\alpha_1 - \frac{2}{3}\alpha_2)^2
        \mathcal{C}\mathcal{D}^2\Big)\Big>.
\eeq
\section{Parameter constraints}
\label{FIVE}
I will now discuss bounds on the $c_n$ that can be derived by
imposing the observational constraints summarized
in~\cite{Foster:2005dk} and by comparing the damping rate
prediction~\eqref{EEB} with measurements of binary pulsar systems.
The four $c_n$ can be reduced to one free parameter by requiring
that the PPN parameters $\alpha_1$ and $\alpha_2$ vanish, and that
the damping rate coincide with that of GR to lowest order. The
theory will then satisfy all solar system based tests; it is not
correct, however, to say that it would pass the binary pulsar
test. This is because the fields inside a neutron star pulsar or
black hole companion are not weak, and strong field corrections to
the quasi-static parameters may arise. Nevertheless, the weak
field results are adequate for small enough $c_n$, as discussed
in~\cite{Foster:2007gr}. Therefore, it is useful to check whether
this curve of ae-theories intersects the region allowed by
positive energy, real frequency, vacuum \v{C}erenkov, and
nucleosynthesis constraints.

\begin{figure}[tb]
\centering \epsfig{file=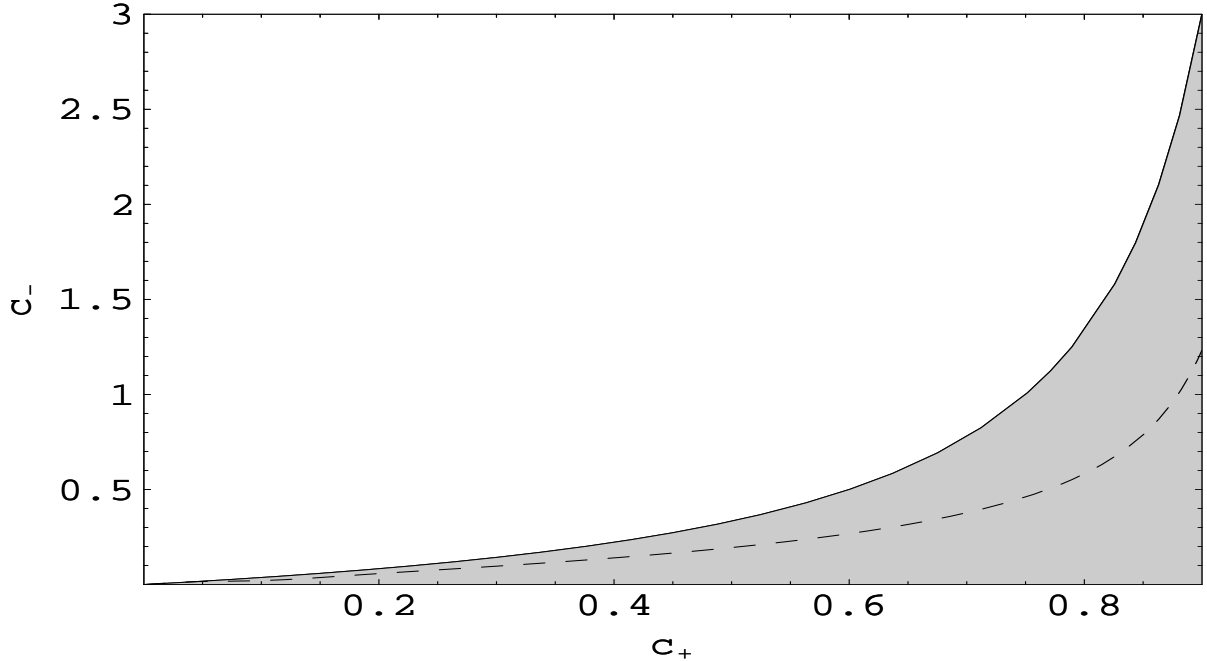, width=\textwidth,
height=3.5in} \caption{Class of allowed ae-theories, if strong
field effects in binary pulsar systems can be ignored. The
four-dimensional $c_n$ space has been restricted to the
$(c_+,c_-)$ plane by setting the PPN parameters $\alpha_1$ and
$\alpha_2$ to zero via the conditions~\eqref{CONDITION5}.  The
shaded region is the region allowed by collected non-binary
constraints, demarcated in~\eqref{REGION}. The dashed curve is the
curve along which binary pulsar tests will be satisfied, assuming
ae-theory weak field expressions. Along both this curve and the
boundary of the allowed region, $c_-\rightarrow \infty$ as $c_+
\rightarrow 1$. The curve remains within the allowed region for
all $c_+$ between 0 and 1. Strong field effects may lead to
system-dependent corrections to the binary pulsar curve for large
$c_n$; however, all such curves will coincide with the weak field
curve for $|c_n| \lesssim (0.01)$ given current observational
uncertainties~\cite{Foster:2006az}.} \label{FIG1}
\end{figure}

The PPN parameters $\alpha_1$~\eqref{alph1} and
$\alpha_2$~\eqref{alph2} for ae-theory were determined
in~\cite{Foster:2005dk}. It was shown that they can be set to
zero, so that all of the ae-theory PPN parameters coincide with
those of GR, with the choices
\beq\label{CONDITION5}
    c_2 = -\frac{2c_1^2 + c_1c_3 - c^2_3}{3c_1},\qquad
    c_4 = -\frac{c_3^2}{c_1}.
\eeq
The positive energy, real frequency, vacuum \v{C}erenkov, and
nucleosynthesis constraints can then be satisfied if $c_1$ and
$c_3$ lie within the region
\beq\label{REGION}
    0 < c_+ < 1,\qquad
    0< c_- < \frac{c_+}{3(1-c_+)}.
\eeq
When $\alpha_1$ and $\alpha_2$ vanish, so does $Z$~\eqref{ZZZ}
hence $\mathcal{B}$~\eqref{BBB5}, and $\Sigma_i$. The fields then
contain only a quadrupole contribution, and the ae-theory damping
rate~\eqref{EEB} will match that of GR when $\mathcal{A} = 1$.
Solving numerically shows that a solution curve exists in
$(c_+,c_-)$ space that intersects the allowed
region~\eqref{REGION} for all positive values of $c_-$; see
Figure~\ref{FIG1}. Thus, there exists a one-parameter family of
ae-theories which satisfy all of the constraints summarized
in~\cite{Foster:2005dk}, and which predict a damping rate
identical in the weak field limit to that of GR.

Observational error allows this curve to be widened into a band.
As explained in the Introduction, the standard method of measuring
radiation damping is to observe the rate of change of orbital
period $\dot{P}$ of a binary
system~\cite{Will:2001mx,Stairs:2003eg}, which will be
proportional to $\dot{\mathcal{E}}$. The smallest relative
observational uncertainty in $\dot{P}$, which equates with the
relative uncertainty in $\dot{\mathcal{E}}$, is of order $0.1\%$
for the Hulse-Taylor binary
B1913+16~\cite{Will:2001mx,Stairs:2003eg}. This uncertainty
permits the band $|\mathcal{A} - 1| \lesssim 10^{-3}$. Numerical
results indicate that at least for small $c_{\pm}$, this band
corresponds roughly to $c_{\pm}$ within about $10^{-3}$ of the
$\mathcal{A} = 1$ curve.
%
%
%
\begin{acknowledgments}
I would like to thank Ted Jacobson for editorial acumen. This
research was supported in part by the NSF under grant PHY-0300710
at the University of Maryland.
\end{acknowledgments}
%
%

%
\end{document}